\documentclass[twoside,twocolumn,9pt]{article}
\usepackage{extsizes}
\usepackage[super,sort&compress,comma]{natbib} 
\usepackage[version=3]{mhchem}
\usepackage[left=1.5cm, right=1.5cm, top=1.785cm, bottom=2.0cm]{geometry}
\usepackage{balance}
\usepackage{times,mathptmx}
\usepackage{sectsty}
\usepackage{graphicx} 
\usepackage{lastpage}
\usepackage[format=plain,justification=justified,singlelinecheck=false,font={stretch=1.125,small,sf},labelfont=bf,labelsep=space]{caption}
\usepackage{float}
\usepackage{fancyhdr}
\usepackage{fnpos}
\usepackage[english]{babel}
\addto{\captionsenglish}{%
  \renewcommand{\refname}{Notes and references}
}
\usepackage{array}
\usepackage{droidsans}
\usepackage{charter}
\usepackage[T1]{fontenc}
\usepackage[usenames,dvipsnames]{xcolor}
\usepackage{setspace}
\usepackage[compact]{titlesec}
\usepackage{hyperref}

\usepackage{epstopdf}

\definecolor{cream}{RGB}{222,217,201}

\begin{document}

\pagestyle{fancy}
\thispagestyle{plain}
\fancypagestyle{plain}{

\fancyhead[C]{\includegraphics[width=18.5cm]{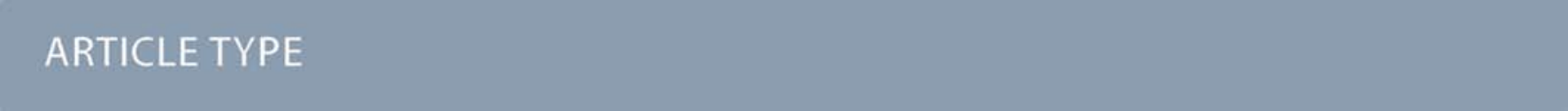}}
\fancyhead[L]{\hspace{0cm}\vspace{1.5cm}\includegraphics[height=30pt]{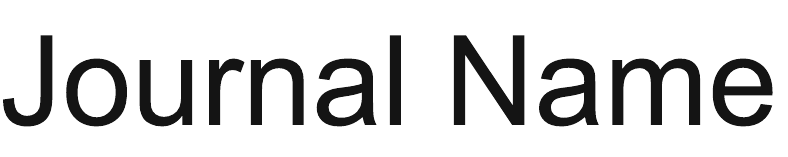}}
\fancyhead[R]{\hspace{0cm}\vspace{1.7cm}\includegraphics[height=55pt]{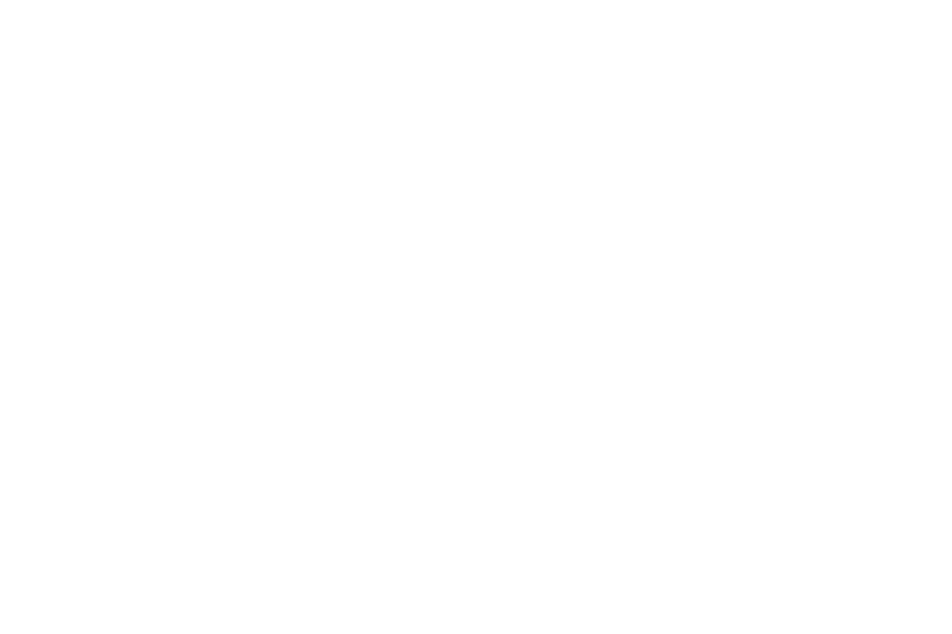}}
\renewcommand{\headrulewidth}{0pt}
}

\makeFNbottom
\makeatletter
\renewcommand\LARGE{\@setfontsize\LARGE{15pt}{17}}
\renewcommand\Large{\@setfontsize\Large{12pt}{14}}
\renewcommand\large{\@setfontsize\large{10pt}{12}}
\renewcommand\footnotesize{\@setfontsize\footnotesize{7pt}{10}}
\makeatother

\makeatother

\renewcommand\refname{}
\newcommand{\fl}[1]{{\color{green}#1}}
\newcommand{\ul}[1]{{\color{red}#1}}
\newcommand{\nb}[1]{{\color{blue}#1}}
\newcommand{\hL}{\hat{L}}
\newcommand{\Tf}{T_{flat}}
\newcommand{\Ts}{T_s}
\newcommand{\Ws}{W_s}
\newcommand{\Uf}{U_{flat}}
\newcommand{\Us}{U_{s}}
\newcommand{\Usi}{U_{s,i}}
\newcommand{\tl}{\theta_{l,i}}

\renewcommand{\thefootnote}{\fnsymbol{footnote}}
\renewcommand\footnoterule{\vspace*{1pt}%
\color{cream}\hrule width 3.5in height 0.4pt \color{black}\vspace*{5pt}} 
\setcounter{secnumdepth}{5}

\makeatletter 
\renewcommand\@biblabel[1]{#1}            
\renewcommand\@makefntext[1]%
{\noindent\makebox[0pt][r]{\@thefnmark\,}#1}
\makeatother 
\renewcommand{\figurename}{\small{Fig.}~}
\sectionfont{\sffamily\Large}
\subsectionfont{\normalsize}
\subsubsectionfont{\bf}
\setstretch{1.125} 
\setlength{\skip\footins}{0.8cm}
\setlength{\footnotesep}{0.25cm}
\setlength{\jot}{10pt}
\titlespacing*{\section}{0pt}{4pt}{4pt}
\titlespacing*{\subsection}{0pt}{15pt}{1pt}

\fancyfoot{}
\fancyfoot[LO,RE]{\vspace{-7.1pt}\includegraphics[height=9pt]{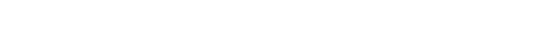}}
\fancyfoot[CO]{\vspace{-7.1pt}\hspace{13.2cm}\includegraphics{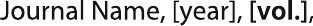}}
\fancyfoot[CE]{\vspace{-7.2pt}\hspace{-14.2cm}\includegraphics{head_foot/RF}}
\fancyfoot[RO]{\footnotesize{\sffamily{1--\pageref{LastPage} ~\textbar  \hspace{2pt}\thepage}}}
\fancyfoot[LE]{\footnotesize{\sffamily{\thepage~\textbar\hspace{3.45cm} 1--\pageref{LastPage}}}}
\fancyhead{}
\renewcommand{\headrulewidth}{0pt} 
\renewcommand{\footrulewidth}{0pt}
\setlength{\arrayrulewidth}{1pt}
\setlength{\columnsep}{6.5mm}
\setlength\bibsep{1pt}

\makeatletter 
\newlength{\figrulesep} 
\setlength{\figrulesep}{0.5\textfloatsep} 

\newcommand{\topfigrule}{\vspace*{-1pt}%
\noindent{\color{cream}\rule[-\figrulesep]{\columnwidth}{1.5pt}} }

\newcommand{\botfigrule}{\vspace*{-2pt}%
\noindent{\color{cream}\rule[\figrulesep]{\columnwidth}{1.5pt}} }

\newcommand{\dblfigrule}{\vspace*{-1pt}%
\noindent{\color{cream}\rule[-\figrulesep]{\textwidth}{1.5pt}} }

\makeatother

\twocolumn[
  \begin{@twocolumnfalse}
\vspace{3cm}
\sffamily
\begin{tabular}{m{4.5cm} p{13.5cm} }

\includegraphics{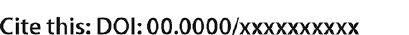} & \noindent\LARGE{\textbf{Droplet leaping governs microstructured surface wetting}} \\
\vspace{0.3cm} & \vspace{0.3cm} \\

 & \noindent\large{Susumu Yada$^{a}$, Shervin Bagheri$^{a}$, Jonas Hansson$^{b}$, Minh Do-Quang$^{a}$, Fredrik Lundell$^{a}$, Wouter van der Wijngaart$^b$ and Gustav Amberg$^{a,c}$} \\

\includegraphics{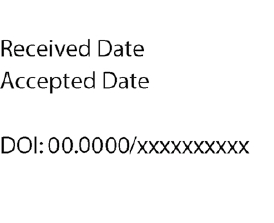} & \noindent\normalsize{Microstructured surfaces that control the direction of liquid transport are not only ubiquitous in nature, but they are also central to technological processes such as fog/water harvesting, oil-water separation, and surface lubrication. However, a fundamental understanding of the initial wetting dynamics of liquids spreading on such surfaces is lacking.
Here, we show that three regimes govern microstructured surface wetting on short time scales: spread, stick, and contact line leaping. The latter involves establishing a new contact line downstream of the wetting front as the liquid leaps over specific sections of the solid surface. 
Experimental and numerical investigations reveal how different regimes emerge in different flow directions during wetting of periodic asymmetrically microstructured surfaces. 
These insights improve our understanding of rapid wetting in droplet impact, splashing, and wetting of vibrating surfaces and may contribute to advances in designing structured surfaces for the mentioned applications.}\\
\end{tabular}
 \end{@twocolumnfalse} \vspace{0.6cm}
]

\renewcommand*\rmdefault{bch}\normalfont\upshape
\rmfamily
\vspace{-1cm}
%
\footnotetext{\textit{$^{a}$~Department of Mechanics, KTH Royal Institute of Technology, 100 44 Stockholm, Sweden.; E-mail: shervin@mech.kth.se}}
\footnotetext{\textit{$^{b}$~Division of Micro and Nanosystems, KTH Royal Institute of Technology, 100 44 Stockholm, Sweden. }}
\footnotetext{\textit{$^{c}$~S{\"o}dertorn University, Stockholm, Sweden. }}
\footnotetext{\dag~Electronic Supplementary Information (ESI) available: See DOI: 00.0000/00000000.}
%
%
\section{Introduction}
Natural surfaces with texture provide organisms the ability to control liquid transport in fascinating ways. 
For example, lotus leaves have hierarchically structured surfaces that ease droplet removal\cite{Liu2017NRM,GUO20071103}. The ``pitcher plant'' {\it Nepenthes alata} has wettable asymmetric microridges on their peristomes that make the surface slippery and aid in catching insects efficiently\cite{Chen2016, Bohn2004}. Ryegrass leaves have inclined protrusions to help to shed off water unidirectionally\cite{Guo2012}. Cacti in deserts gather water through thorns with surface structures that collect fog effectively\cite{Ju2012}. 

Understanding how liquids act on non-smooth surfaces has also become increasingly important in technological processes, such as printing, coating, adhesion, mixing, and sorting\cite{Cohen1}.  In particular, natural surfaces have inspired the fabrication of micro- and nanoscopic topographies where asymmetric structures are introduced to enable capillary-driven directional liquid transport\cite{Chu2010,Holmes2015, Malvadkar2010}. Chu \textit{et.~al.}\cite{Chu2010} first demonstrated that droplets can spread unidirectionally on arrays of asymmetric nanorods. 
Various applications of asymmetric-structured surfaces have been reported since, including mixing in microreactors\cite{Lin2018}, oil-water separation\cite{Li2016}, and transport in microfluidic channels\cite{Zhang2017}.

The spreading of a liquid droplet over a textured surface has thus far been analyzed in the slow late spreading regime ($\approx$ 100 ms to few seconds)\cite{Bonn2009, Tanner1979, Chu2010, Lin2018,Yu2018}.
The details and mechanisms of such spreading have been described in terms of contact-line pinning\cite{Yu2018, Chen2016}, Laplace pressure\cite{Yu2018, Si2018, Lin2018}, and gradients in surface energy\cite{Shastry2006, Liu2017}. Also, numerical simulations have provided valuable insights into the spreading mechanisms in the late regime\cite{Cavalli2013, Chamakos2016}. However, the rapid wetting that precedes this late spreading is not well understood. Rapid wetting is crucial for droplet impact on solid surfaces\cite{Li2016_Janus, Agapov2014b}, or in situations where time scales are imposed externally, such as wetting phenomena in the presence of vibrations\cite{Habibi2016}.

Here, we theoretically, experimentally and numerically investigate rapid wetting on microstructures. 
We consider the dynamics of the leading contact line on slanted ridges in a two dimensional plane.
%
We elucidate how the contact line follows the microstructures and explain and predict the effect of the geometry on the spreading speed.
\begin{figure}[t!]
  \begin{center}    \includegraphics[width=0.47\textwidth]{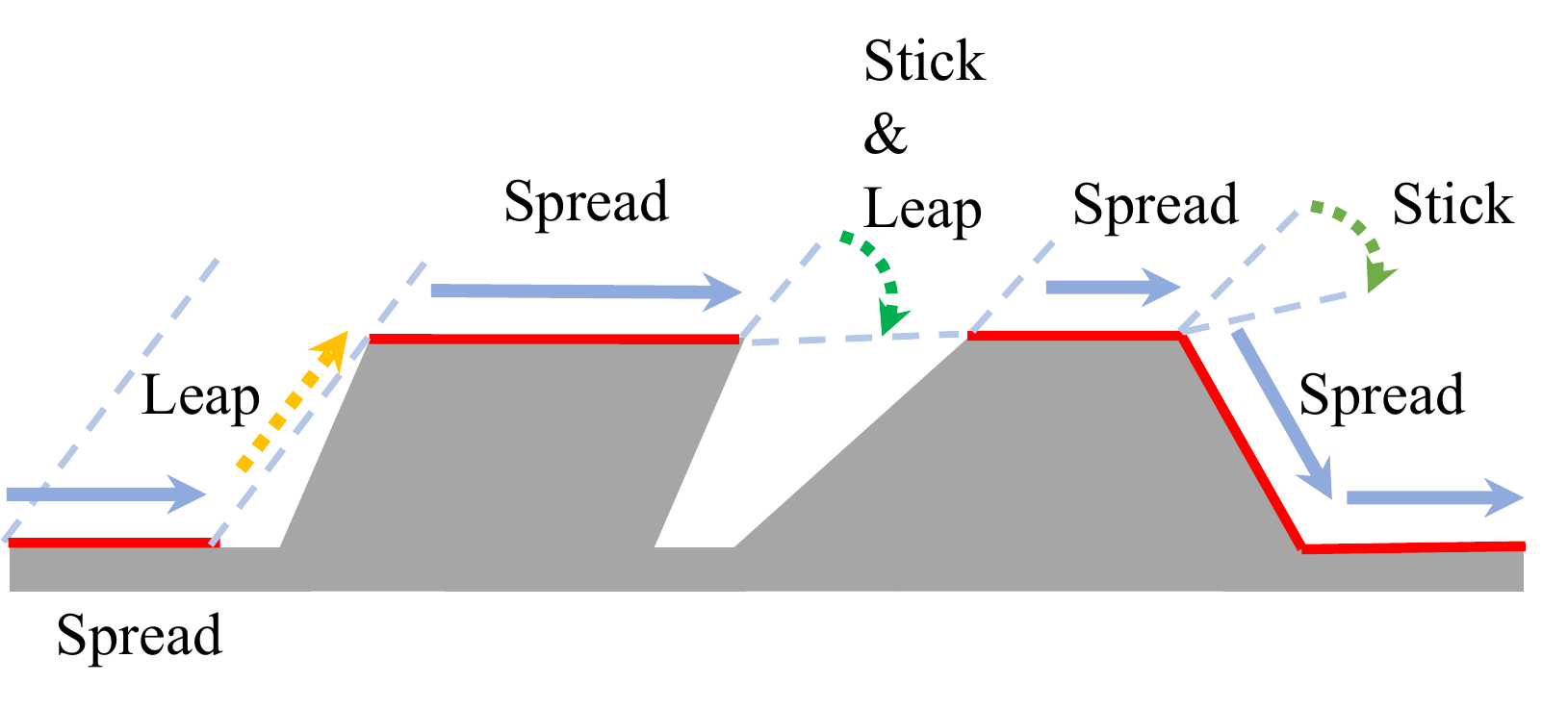}
  \end{center}
  \vspace{-0.2cm}
  \caption{
  Spreading regimes during wetting of microstructured surfaces.
  Cross section of a hypothetical structured surface (grey) as a liquid front moves from left to right. Dashed blue lines mark the instantaneous position of the liquid-air interface and blue arrows indicate the movement direction of the interface. The surface that is wetted by the liquid is marked with red. The contact line spreads on flat and downward-sloping sections, whereas it may stick and/or leap at sections with corners.
  %
  }
\label{fig:map:a}
\end{figure}
We hypothesize that three wetting regimes determine the contact line velocity and that wetting can be understood from the succession of these regimes as the contact line travels over the surface (Fig.~\ref{fig:map:a}). The first regime, called {\it spread}, is the movement of the contact line on a homogeneous surface with a velocity determined by surface chemistry and liquid-gas properties\cite{Vo2018,Do-Quang2015}. The second regime, called {\it stick}, refers to the pinning of the contact line on surface corners, resulting in a temporarily stationary contact line\cite{Mori1982}. Both spread and stick regimes are well understood\cite{Bird2008,Tanner1979,Mori1982} in contrast to the third phenomenon we observed and that we call {\it leap}. Contact line leaping involves the establishment of a new contact line downstream of the wetting front when the liquid leaps over specific sections of the solid surface, trapping gas between the solid and the liquid. As the surface wets, the wetting regimes can follow each other in any order, as illustrated in Fig.~\ref{fig:map:a}.

\section{Experiments and methods}
\subsection{Experimental set-up}
Spontaneous droplet spreading is recorded with a high-speed camera (DANTEK dynamics Speedsense M) at a frame rate of 52,044 per second. A water droplet develops from a needle with outer diameter of 0.31 mm (Hamilton, gauge 30, point style 3) and starts to spread spontaneously immediately after it touches a substrate. Water is pumped by a syringe pump (Cetoni, neMESYS 1000N) at a flow rate 0.04 $\mu$l/s. The flow rate is so small that a static state is assumed before the droplet touches the surface. The initial radius of the droplet $R_0$ is determined by the distance between the needle and the surface and fixed to 0.4 mm. The scope is composed of 128 $\times$128 pixels, which leads to the spatial resolution $\approx$8$\mu$m. As seen in Fig.~\ref{fig:spread}(c), owning to the limited spatial resolution, the observed droplet has finite spreading radius even before it actually touches the substrate. In order to estimate the initial time $t_{0}$ in which the droplet starts to spread, a power-law curve $r= C(t-t_{0})^{\gamma} $ is fitted to the time history of observed spreading radius where $C$, $\gamma$, and $t_0$ are scalar parameters. The initial time is then determined by shifting the original spreading time by $t_0$. 

\subsection{Sample preparation}
The microstructured surfaces are made from Ostemer 220 (Mercene Labs, Stockholm, Sweden), a UV-curing Off-Stoichiometry-Thiol-Ene (OSTE) resin, with excellent lithographic patterning, previously used to create complex slanted structures\cite{Hansson2016}. 
The samples are manufactured as follows. First, a flat Ostemer layer is manufactured on a smooth plastic film. Second, slanted Ostemer ridges are patterned on the base layer by exposing slanted UV light through a patterned photomask. After development in Aceton, a surface modification is performed to hydrophilize the surface (equilibrium contact angle  $\theta_{e} \approx 50^\circ$ on a flat substrate) in order to achieve partial wetting. Surface structures are characterized with scanning electron microscopy  as shown in Supplementary Fig.~1. The inclination of the ridges $\beta$ is 60$^\circ$ for all the structures. The samples used in this work are labeled as ($W$, $P$) where $W$ and $P$ are the width and pitch of the ridge in micrometers, respectively, and listed in Table 1.


\begin{table}[h]
  \caption{Geometry and passage times for different surface structures investigated experimentally. The sample with bold font is also investigated numerically. The numerical values given for width ($W$), pitch ($P$) and height ($H$) are in micrometers. The last two columns show the passage time ratio with and against inclination estimated from experiments. }
\begin{tabular}{lcccll}
     \hline
     Label & $W$ & $P$ & $H$ & $S_{against}$ & $S_{with}$ \\
    \hline
     (10, 15) & 10 & 15 & 14 & 1.39 $\pm$ 0.10 & 1.21 $\pm$ 0.06 \\
     (10, 20) & 10 & 20 & 14 & 1.61 $\pm$ 0.15 & 1.30 $\pm$ 0.11 \\
     (10, 30) & 10 & 30 & 14 & 3.07 $\pm$ 0.50 & 4.66 $\pm$ 0.71 \\
     (15, 23) & 15 & 23 & 14  & 1.30 $\pm$ 0.12 & 1.13 $\pm$ 0.07 \\
     (15, 30) & 15 & 30 & 14 & 1.43 $\pm$ 0.11 & 4.12 $\pm$ 0.58\\
     (15, 45) & 15 & 45 & 14 & 3.11 $\pm$ 0.67 & 6.3 $\pm$ 1.4\\
     (20, 30) & 20 & 30 & 15 & 1.99 $\pm$ 0.33 & 5.19 $\pm$ 0.84 \\
     (20, 40) & 20 & 40 & 17 & 3.77 $\pm$ 0.44 & 5.5 $\pm$ 1.7 \\
     \textbf{(20, 60)} & 20 & 60 & 17 & 1.80 $\pm$ 0.36 & 5.06 $\pm$ 0.62 \\
     \hline
  \end{tabular}
\label{tab1}
  \end{table}

\subsection{Navier-Stokes-Cahn-Hillard equations}
The Navier-Stokes equations combined with the phase-field approach (Cahn-Hilliard equation) are solved to simulate the droplet wetting on a comparable geometry to the experiments. The incompressible Navier-Stokes equations are given by
 \begin{equation}
	\rho(C) \frac{D\textit{\textbf{u}}}{Dt} = \frac{1}{Re} \nabla p+ \frac{1}{Re} \nabla \mu(C) (\nabla \textit{\textbf{u}}+\nabla^T \textit{\textbf{u}})-\frac{C \nabla \phi(C)}{Ca\cdot Cn \cdot Re},
\end{equation}		
\begin{equation}		
	\nabla \cdot \textit{\textbf{u}}=0.			
\end{equation}					
The non-dimensional numbers characterizing the system are the capillary number $Ca$=$\mu U /\sigma$, the Reynolds number $Re$=$\mu UL/\rho$, and the Cahn number $Cn$=$\epsilon/L$. Here, $\rho$ and $\mu$ are the density and viscosity of the liquid phase, $\sigma$ is the surface tension of the liquid, and $\epsilon$ is the  width of the liquid-gas interface. Moreover, $U$and $L$ are the characteristic velocity and length of the system, respectively. The scalar $C$ is the phase field variable, where $C =1$ represents the liquid phase, and $C  =-1$ the vapor phase. The effect of gravity is negligible since the Bond number $Bo$=$\rho gL^{2}/\sigma$ is small ($\approx$0.03 for $L$=0.4 mm).

The Cahn-Hilliard equation is given by
 \begin{equation}
	 \frac{DC}{Dt} = \frac{1}{Pe} \nabla^{2}  \phi(C),
\end{equation}		
where $\phi$ is the chemical potential of the system defined as $\phi=\Psi^{\prime}(C)-Cn \nabla^2 C$. The Peclet number is defined as $Pe=UL/D$ where $D$ is a mass diffusivity. Here, $\Psi(C)= (C+1)^{2}(C-1)^{2}/4$  is the double well function, where the minimum represents the stable phases for gas ($C =-1$) and liquid ($C =1$). The boundary condition for $C$ on a solid surface is given by\cite{CarlsonPoF2009, CarlsonPRE2012}
\begin{equation}
	-\epsilon \mu_{f}  \frac{\partial C}{\partial t} =\epsilon \sigma \nabla C \cdot \textit{\textbf{n}} - \sigma {\rm cos} (\theta _{e})g^{\prime} (C),
\end{equation}		
where \(\theta_{e}\) is equilibrium contact angle and $g(C)= 0.5+0.75C-0.25C^3$ is a polynomial which rapidly shifts from $0$ (vapor phase $C =-1$) to $1$ (liquid phase $C =1$). The line friction parameter $\mu_{f}$ is associated with molecular-origin energy dissipation at the moving contact line.

\begin{figure*}[t!]
  \begin{center}    
  \includegraphics[width=0.7\textwidth]{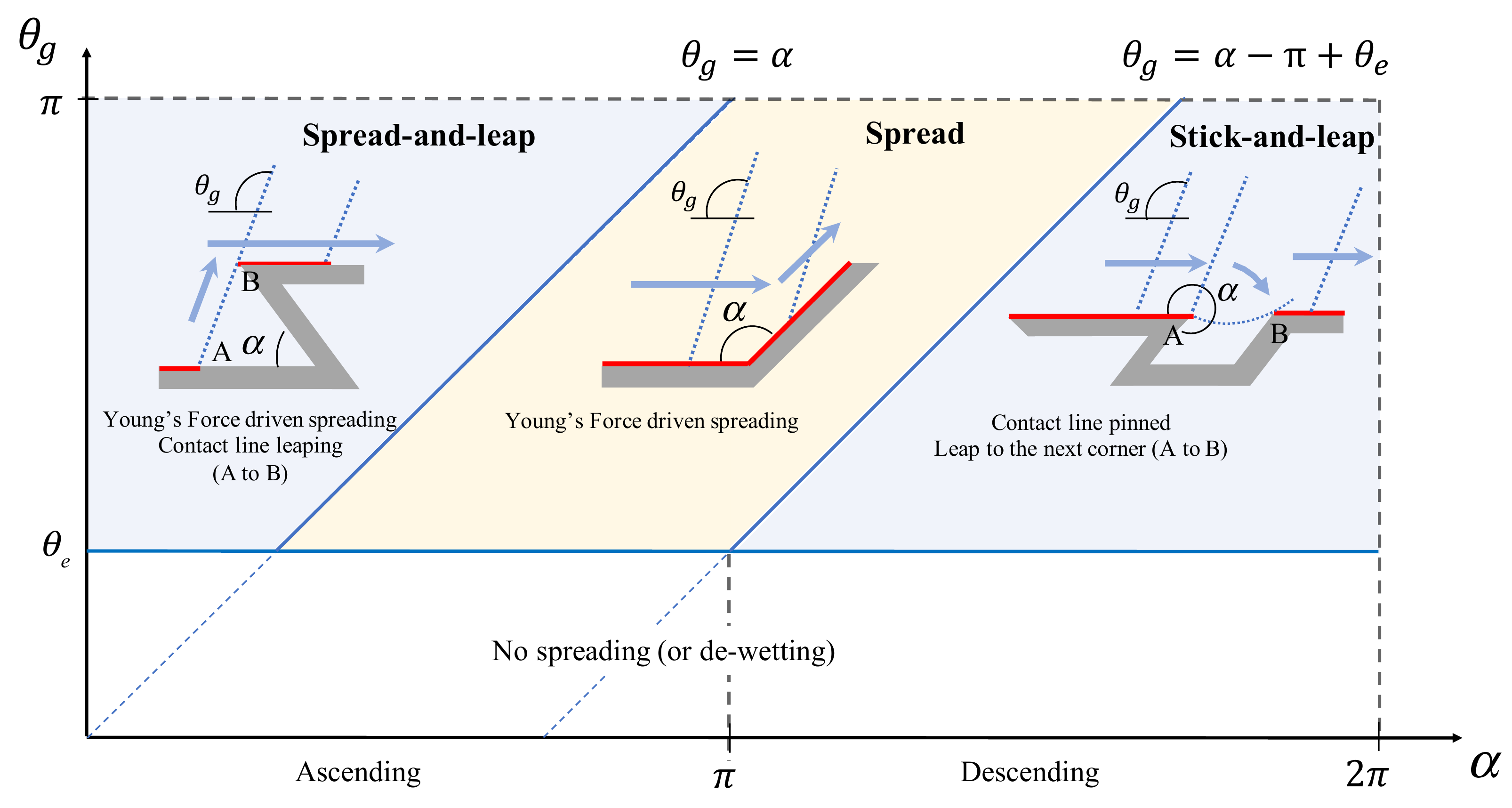}
  \end{center}
  \vspace{-0.2cm}
  \caption{
  %
  %
  The different combinations of spread, stick and leap depends on the globally observed apparent dynamic contact angle ($\theta_g$) and the corner angle ($\alpha$).  A liquid front moving due to Young's force ascends from a valley when $\alpha<\pi$ and descends from a rise when  $\alpha>\pi$. As $\alpha$ increases from acute to obtuse and then to reflex, a liquid front with $\theta_g$ may proceed by spread-and-leap, spread only and stick-leap, respectively.
  }
\label{fig:map:b}
\end{figure*}

\subsubsection{Numerical simulations and parameters}
The Navier-Stokes-Cahn-Hilliard equations are solved using in-house software called FemLego\cite{CarlsonPoF2009}.
The physical properties are chosen to be comparable to the experiments. The characteristic density $\rho$ and viscosity $\mu$ are set to the values of pure water (0.992 kg/m$^{3}$ and 0.997 Pa$\cdot$s respectively), and the length scale $L$ is set to 0.4 mm to match with the initial radius of the droplet in the experiments, which leads to $Re$=29200. Surface tension $\sigma$ is fixed to 0.073 N/m, which gives the capillary speed $\sigma/\mu$=73 m/s employed as the characteristic velocity $U$, and leads to $Ca$=1.
The mass diffusivity $D$ in the Peclet number is set to 5.7$\times$ 10$^{-6}$ m$^{2}$/s, which leads to $Pe$=5120. The choice of Peclet number does not influence the results. The line friction parameter $\mu_{f}$ is obtained by fitting\cite{CarlsonPRE2012, Vo2018} the simulated spreading radius with the experiments on a smooth substrate and found to be $\mu_{f}$=0.10 Pa$\cdot$s (See Supplementary Fig.~2). The interface thickness $\epsilon$ is set to 2 $\mu$m, and results in $Cn$=5$\times$10$^{-3}$. The interface is thicker than the actual physical interface
but it is sufficiently thin compared to the structures in order not to influence the simulated results. 

\section{Wetting mechanism} 
\subsection{Wetting regimes}
\label{sec:regime}
For wetting of structured surfaces, the combination of spread, stick and leap that ultimately determines the contact-line speed depends on the relation between the globally observed apparent dynamic contact angle $\theta_g$, and the corner angle of the surface, $\alpha$ (Fig.~\ref{fig:map:b}). The former angle is the apparent angle between the solid and the liquid-gas interface measured from the wetted side in the vicinity of the contact line.  The corner angle defines the local corner of the surface structure that is approached by a moving contact line (see insets in Fig.~\ref{fig:map:b}). For the rapid wetting considered here, contact line movement is driven by Young's force per unit length of a contact line $F_Y \sim \sigma (\cos \theta_e -\cos \theta_g)$.
During spontaneous spreading, we have  $\theta_e<\theta_g<\pi$.	
 
Three different spreading mechanisms can be distinguished (Fig.~\ref{fig:map:b}); (i) a spread-and-leap behavior of the contact line when $\theta_g>\alpha$; (ii) a stick-and-leap behavior of the contact line when $\theta_g < \alpha-\pi+\theta_e$, and; (iii) continuous spreading driven by $F_Y$ for intermediate corner angles $\alpha-\pi+\theta_e<\theta_g<\alpha $. 
For spread-and-leap, the moving contact line leaps from the valley to the nearest ridge, leaving behind some dry surface.  For stick-and-leap, the contact line is held pinned\cite{Quere2008} at the corner while the liquid-air interface above the pinning site bulges due to inertial forces. 
If there exists a rise of the surface sufficiently nearby, the interface eventually makes contact with the next rise in the texture, resulting in a leap that leaves dry the entire valley between the pinning position and the new contact line.  For intermediate angles, where no leaping occurs and the liquid wets the dry solid, the contact line speed varies with the slope of the surface. In particular, a reduced speed is expected in downward-sloping sections of the surface, compared to flat sections, due to a reduced Young's force $F_Y$.

\begin{figure*}[h!]
  \centering
    \includegraphics[width=0.59\textwidth]{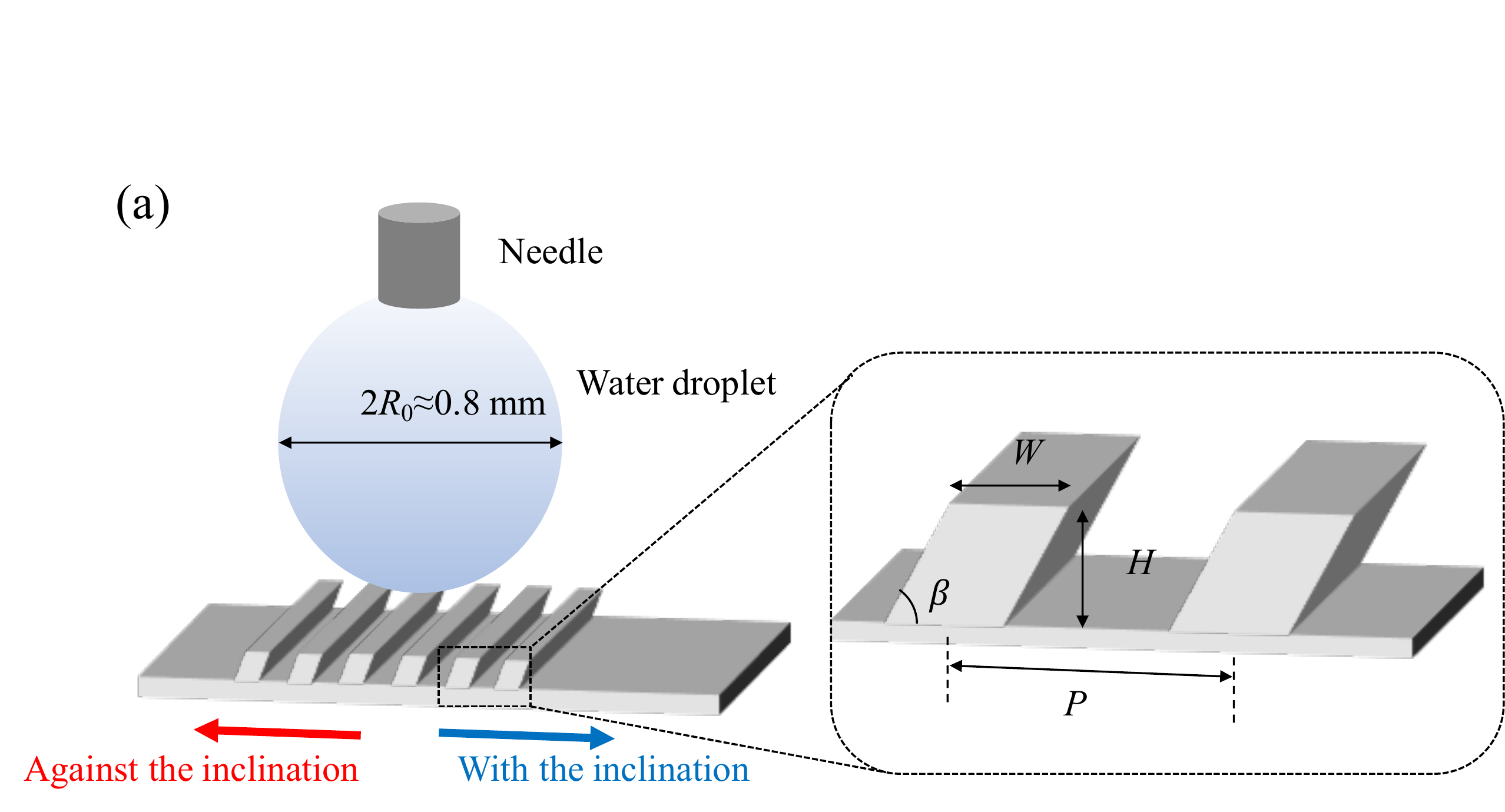}\vspace{0.8cm}
    \includegraphics[width=0.69\textwidth]{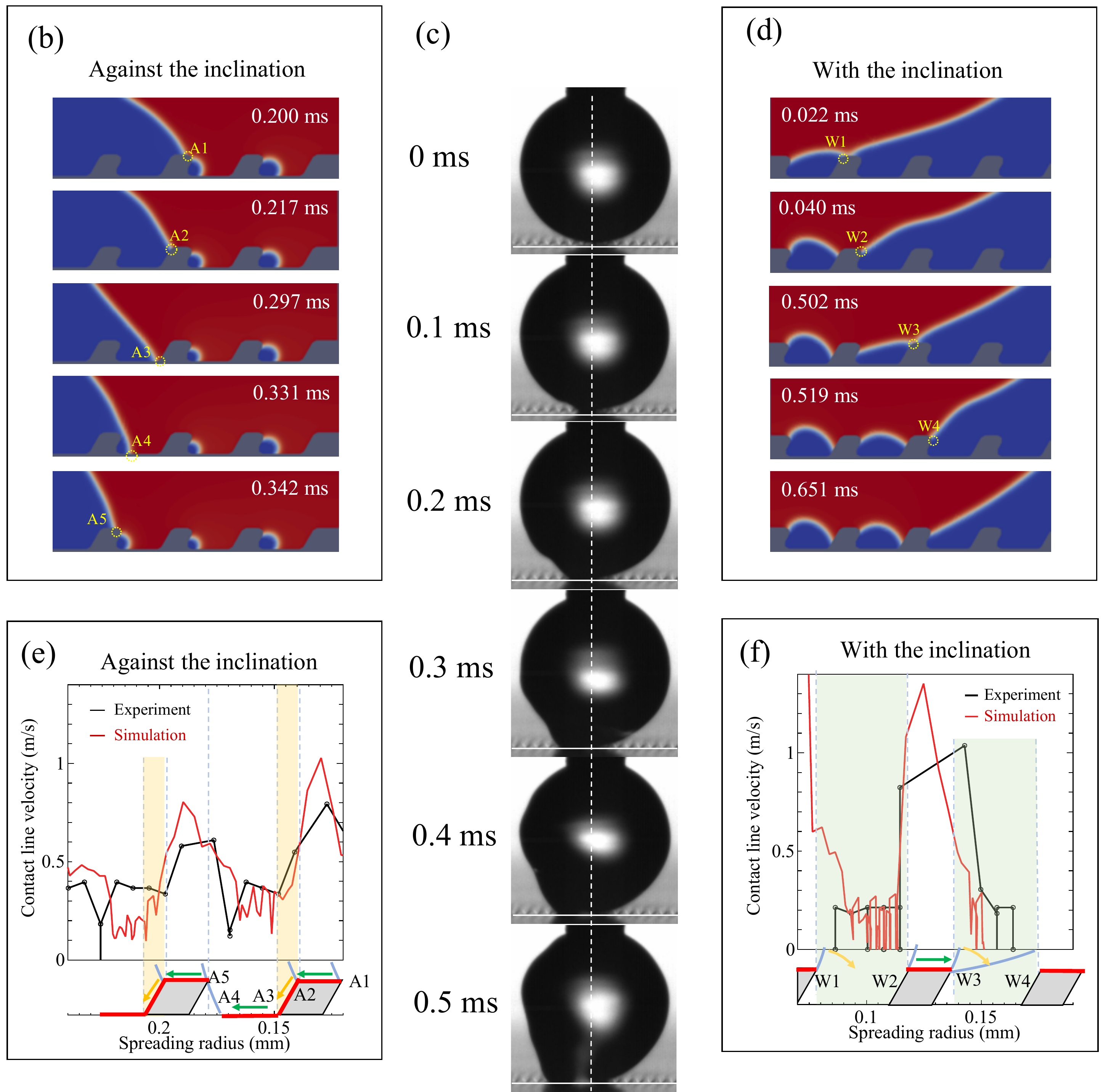}
  \vspace{-0.2cm}
  \caption{
  Liquid spreading on asymmetric periodically microstructured surfaces.
  (a) Schematic illustration of droplet spreading experiments.
  (c) Video frames of a water droplet at different points in time after contact with the microstructured surface.  The spreading distance and droplet shape become increasingly asymmetric. The drop travels faster in the left direction (against the inclination) compared to the right direction (with the inclination).
  (b,d)  Instantaneous numerical snapshots of the interface (white) between gas (blue) and liquid (red) in the two different directions. The points A1-A5 and W1-W4 mark different fixed positions on the surface. 
  (e,f) The measured and simulated contact line velocity along the asymmetric microstructure in the two directions. The points A1-A5 and W1-W4 in (b,d) and (e,f) mark the same fixed positions. }
\label{fig:spread}
\end{figure*}
%
\subsection{Experimental and numerical observations} 
We experimentally investigated wetting of asymmetric ridges by depositing a water drop of radius $0.4$ mm on the surfaces and recording the first $1.2$ ms of wetting using a high-speed video camera (Fig.~\ref{fig:spread}a). We also numerically simulated spreading of droplets on axisymmetric textured surfaces with the same height, pitch and angle as in our experiments using a phase-field approach\cite{CarlsonPoF2009, Wang2015}. 

Figure \ref{fig:spread}(c) shows a time sequence of a water drop spreading on a surface with $W=20$ $\mu$m, $P= 60$ $\mu$m, $H=17$ $\mu$m. The experiments reveal the asymmetric evolution of the droplet spreading on the structure; the contact line travels faster against the inclination (passing five ridges in $0.5$ ms) than with the inclination (passing two ridges in $0.5$ ms). Our numerical simulations reproduce the experimental droplet shapes and the apparent contact line velocity with respect to the spreading radius (see Supplementary Fig.~3 and Supplementary Movie 1). This agreement between experiment and numerical model allows us to rely on the simulations for understanding the detailed contact line movement across the textured surface. 

The difference between the wetting dynamics in the two directions results from the difference in wetting regimes. Figure \ref{fig:spread}(b) shows a time sequence of the liquid-gas interface as the droplet travels against the direction of inclination over one ridge. Figure \ref{fig:spread}(e) shows how the corresponding contact line velocity varies during the same time interval. 
We observe a fast spread on ridge tips (A1$\rightarrow$A2), slower spread while descending into the valley (A2$\rightarrow$A3), and finally an increase in speed with a spread-and-leap to the next ridge (A3$\rightarrow$A4$\rightarrow$A5). 
During the process of moving from A1 to A5, the simulation-predicted apparent dynamic contact angle is in the range of $\theta_g=140^\circ$ to $\theta_g=120^\circ$ (Supplementary Fig.~4), whereas the corner angles encountered by the moving contact line have the values $\alpha=180^\circ$(A1), $240^\circ$(A2), $120^\circ$(A3), and $60^\circ$(A4-A5). 
Thus, our numerical observations conform with the kinematic map in Fig.~\ref{fig:map:b}, explaining the mechanisms behind the peaks and troughs of the contact line speed. 

Figure \ref{fig:spread}(d) shows spreading in the direction with the inclination over one periodic structure. The corresponding contact line speed (Fig.~\ref{fig:spread}f) has a distinct slow-fast-slow spreading velocity going from point W1 to point W4. This velocity profile corresponds to stick-and-leap (W1$\rightarrow$W2) at the corner W1, followed by fast spread on the flat top of the ridge (W2$\rightarrow$W3) and then again followed by stick-and-leap (W3$\rightarrow$W4). As the liquid front moves from W1 to W3, the apparent dynamic contact angle varies from $\theta_g=160^\circ$ to $\theta_g=130^\circ$ (Supplementary Fig.~4).
The stick-and-leap follows from $\theta_g<\alpha-\pi+\theta_e=170^\circ$ at the corners W1 and W3.

Further insight into the contact-line speed can be gained by characterizing the driving and resisting forces during initial wetting. The Young's force driving the contact line can be balanced by an inertial force $F_I\sim\rho \Uf^2 R$ or a contact line friction force $F_f\sim\mu_f \Uf$, where $\Uf$ is a characteristic contact line speed on a smooth surface. The contact line friction is related to non-hydrodynamic energy dissipation at the contact line, connected to molecular scale processes represented by the line friction parameter, $\mu_f$\cite{Johansson2017, CarlsonPoF2009}. The viscous resistance associated with bulk liquid viscosity ($\mu$) can be neglected in the early stages of wetting since $\mu_f\gg \mu$\cite{Do-Quang2015}.

The Ohnesorge number based on the line friction parameter, $Oh_f = \mu_f/\sqrt{\sigma\rho R_0}$, expresses the relative importance of line friction and fluid inertia.  For a water droplet of radius $R_0=0.4$ mm, $Oh_f \approx 0.6$, where we computed the contact line friction to $\mu_f=0.10$ Pa$\cdot$s by combining experiments and numerical simulations (see Section 2 and Supplementary Fig.~2 for details). Since $Oh_f$ is of order one, both inertia and contact line friction may be involved in resisting the spreading. However, they play very different roles during spread-and-leap compared to stick-and-leap. As we will show in the following, in the former regime line friction dominates, whereas, in the latter, inertial forces determine the contact-line speed.

\begin{figure*}[t!]
  \begin{center}
       \includegraphics[width=0.32\textwidth]{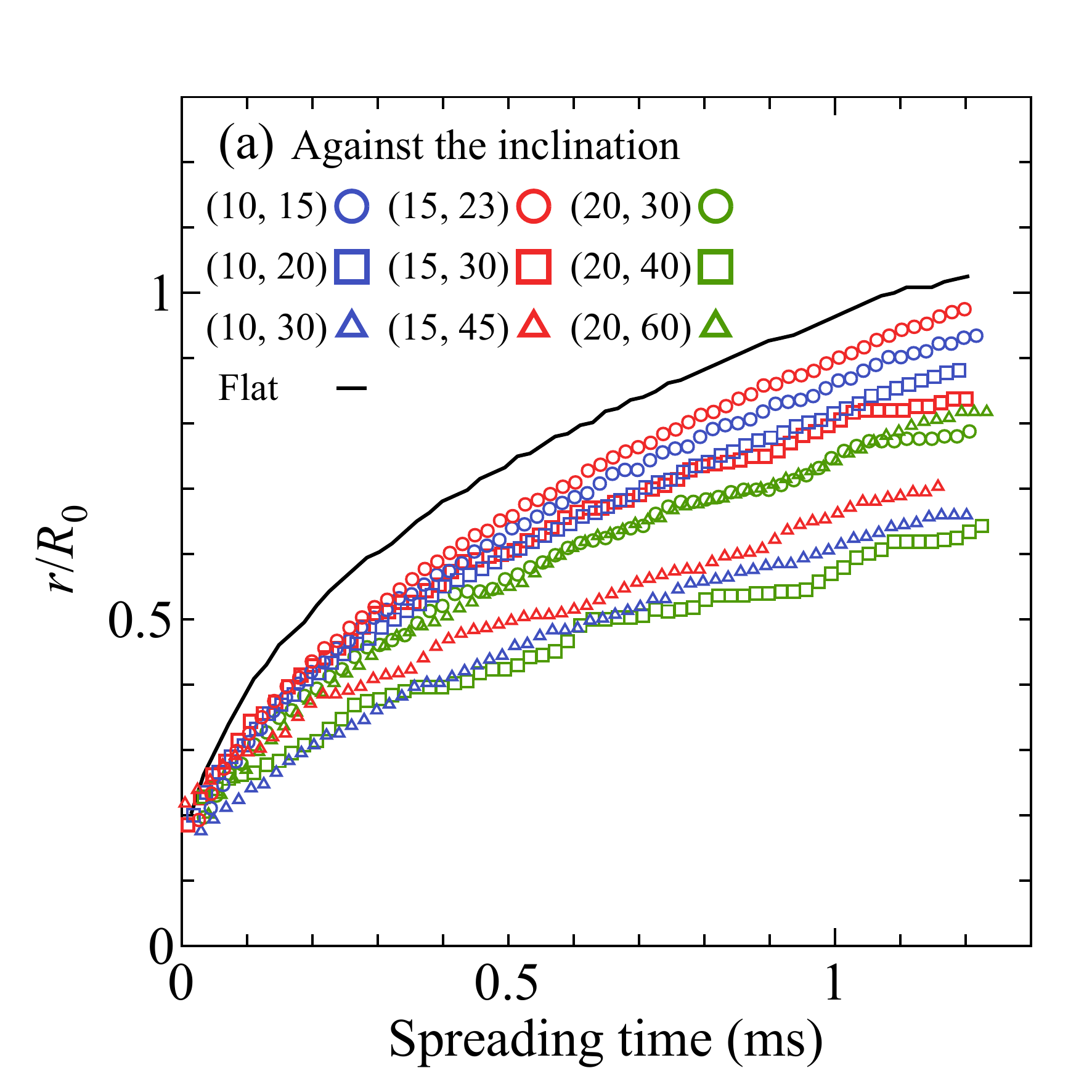}
       \includegraphics[width=0.32\textwidth]{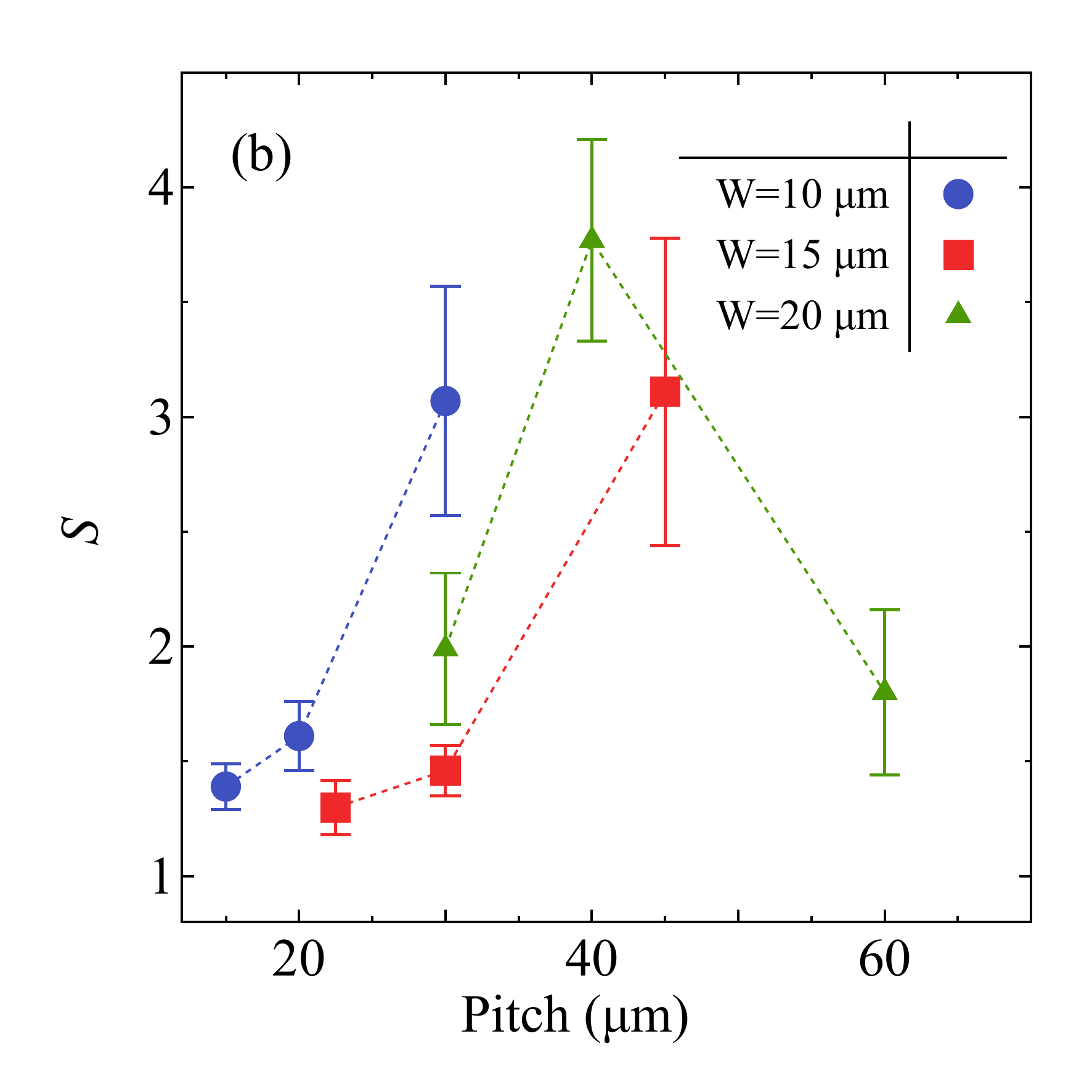} \\
       \includegraphics[width=0.32\textwidth]{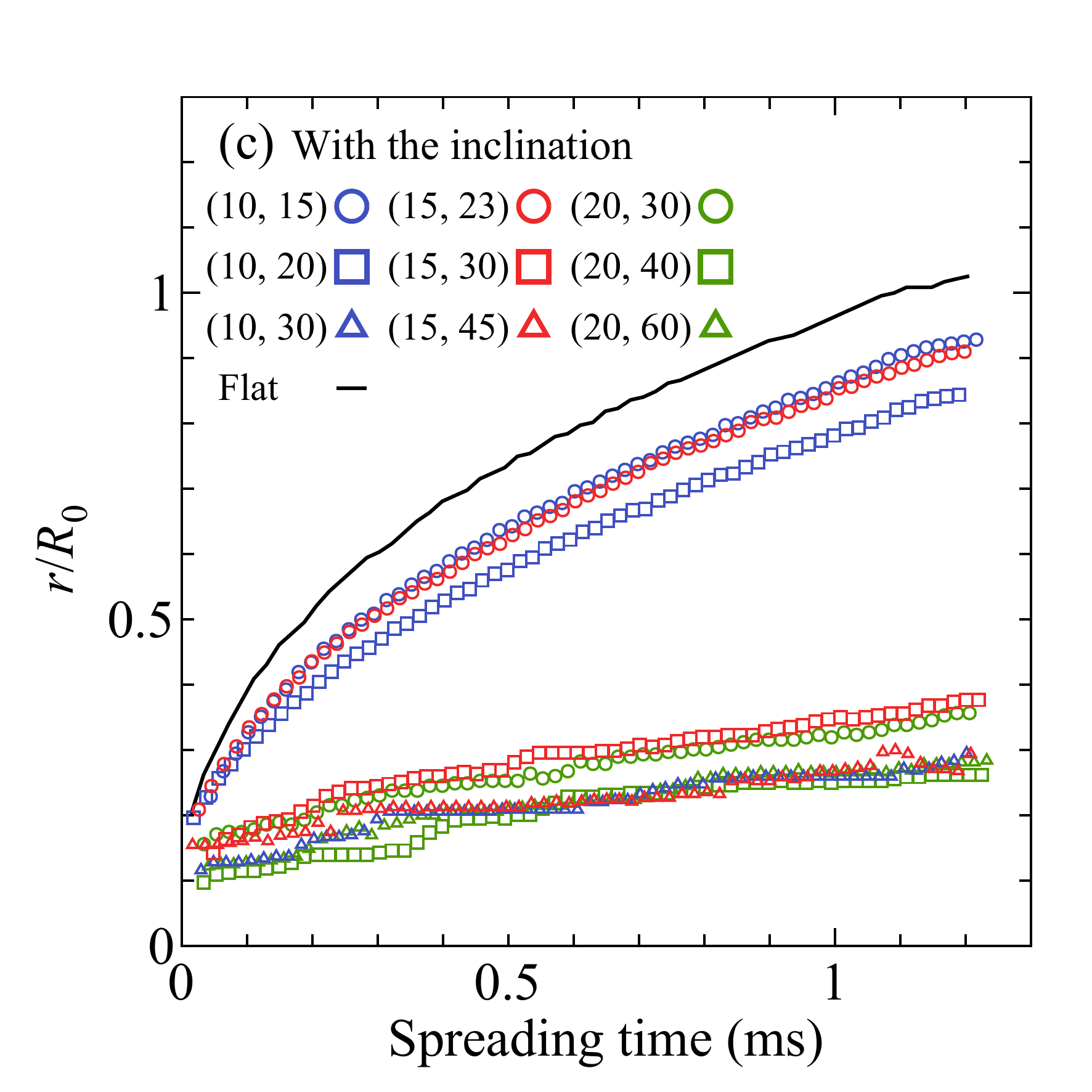}
       \includegraphics[width=0.32\textwidth]{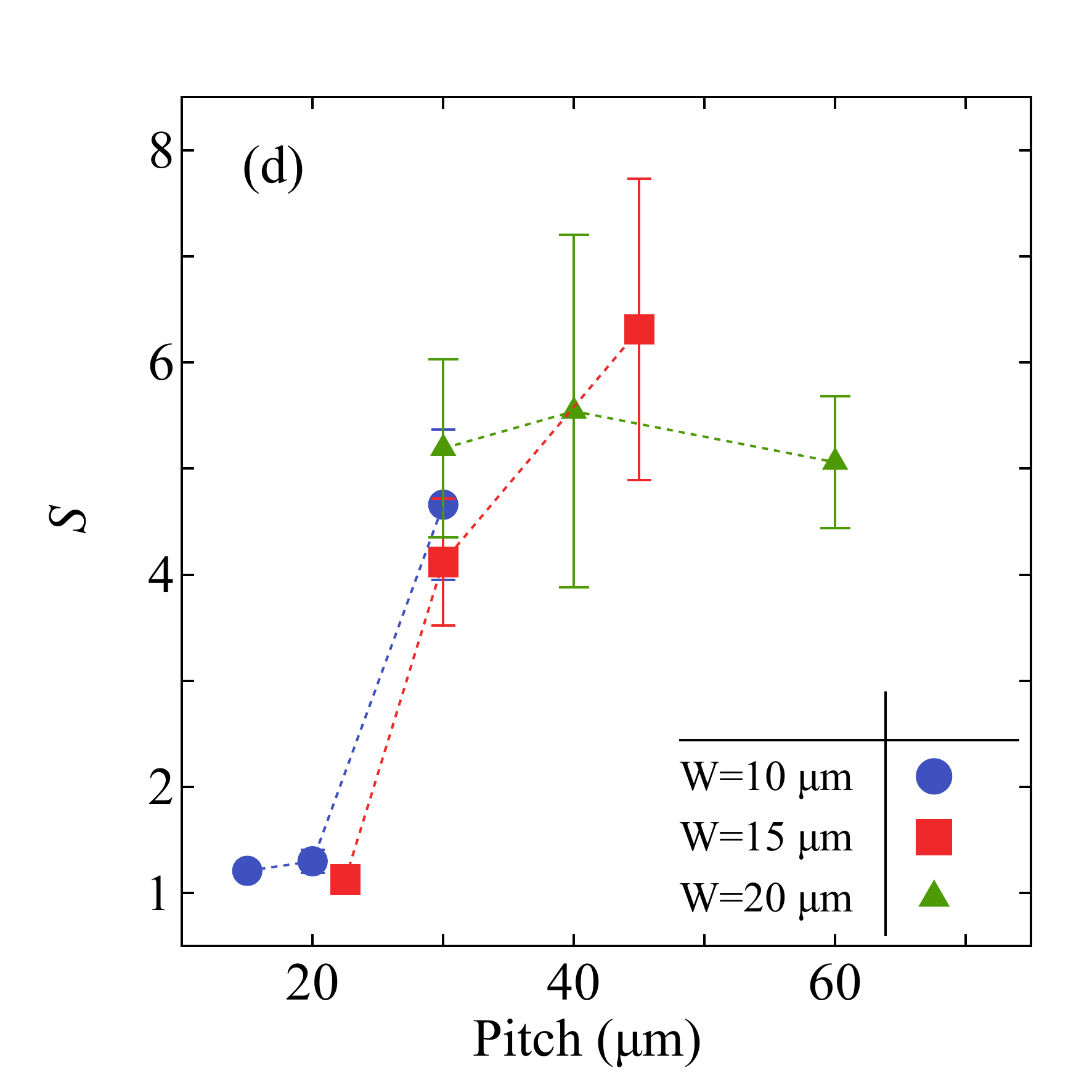}
  \end{center}
  \vspace{-0.2cm}
  \caption{
  (a) The spreading radius versus time during wetting for microstructures of widths $W=10$ $\mu$m (blue), $W=15$ $\mu$m (red) and $W=20$ $\mu$m (green) at three different pitches in the direction against the inclination. Solid line represent spreading radius of smooth surface. The spreading radii are averaged over 5 experimental realizations.  
  (b) The passage time ratio $S$ in the direction against the inclination as a function of surface pitch. Symbols are estimated from wetting experiments shown in (a). 
  (c)  The spreading radius in the direction with the inclination for width $W=10$ $\mu$m (blue), $W=15$ $\mu$m (red) and $W=20$ $\mu$m (green) at three different pitches. 
  (d) The passage time ratio $S$ in the direction with the inclination as a function of surface pitch. Symbols are estimated from wetting experiments shown in (c). Error bars in (b) and (d) indicate standard deviations.}
\label{fig:pitch}
\end{figure*}

\vspace{0.4cm}
\section{Influence of the pitch on spreading speed}
\subsection{Passage time ratio}
Figure \ref{fig:pitch}(a) shows the experimental spreading radius evolution in the direction against the inclination for surfaces with different width ($W$) and pitch ($P$). We observe that the spread on structured surfaces is always slower compared to a smooth surface (solid line).
To characterize the spread-and-leap regime quantitatively, we define the passage time ratio,
\begin{equation}
S = \frac{\Ts}{\Tf},
\label{eq:Sa}
\end{equation}
where $\Ts$ and $\Tf$ correspond to the liquid front travel time over the distance $P$ on structured surfaces and smooth surfaces, respectively. Large values of $S$ thus indicate slow spreading on structured surfaces compared to smooth surfaces. Figure \ref{fig:pitch}(b) shows that passage time $S$ increases nearly by a factor three when the pitch is increased from $15$ $\mu$m to $30$ $\mu$m for W=10 $\mu$m.  A similar increase in $S$ is also observed for W=15 $\mu$m. However, for wider structures W=20 $\mu$m (Fig.~\ref{fig:pitch}b), the same rapid increase is observed  when the pitch is increased from $30$ $\mu$m to $40$ $\mu$m; but the passage time reduces for very large pitch ($60$ $\mu$m).
%

Figure \ref{fig:pitch}(c) shows how the experimental spreading radius evolves in the direction with the inclination for surfaces with a different pitch. Compared to the smooth surface, we observe a slow spreading, which rapidly decreases with the pitch.  
From the corresponding passage times $S$ (symbols in Fig.~\ref{fig:pitch}(d)), $S$ increases monotonically for small $P$ and saturates to $S \sim 6$ for large $P$. 
We note that the travel time can be a factor five higher compared to a smooth surface, and thus also  significantly larger compared to the spread-and-leap regime considered in the previous section (c.f.~Fig.~\ref{fig:pitch}(b, d)).

\begin{figure}[t!]
  \begin{center}  
  \includegraphics[width=0.35\textwidth]{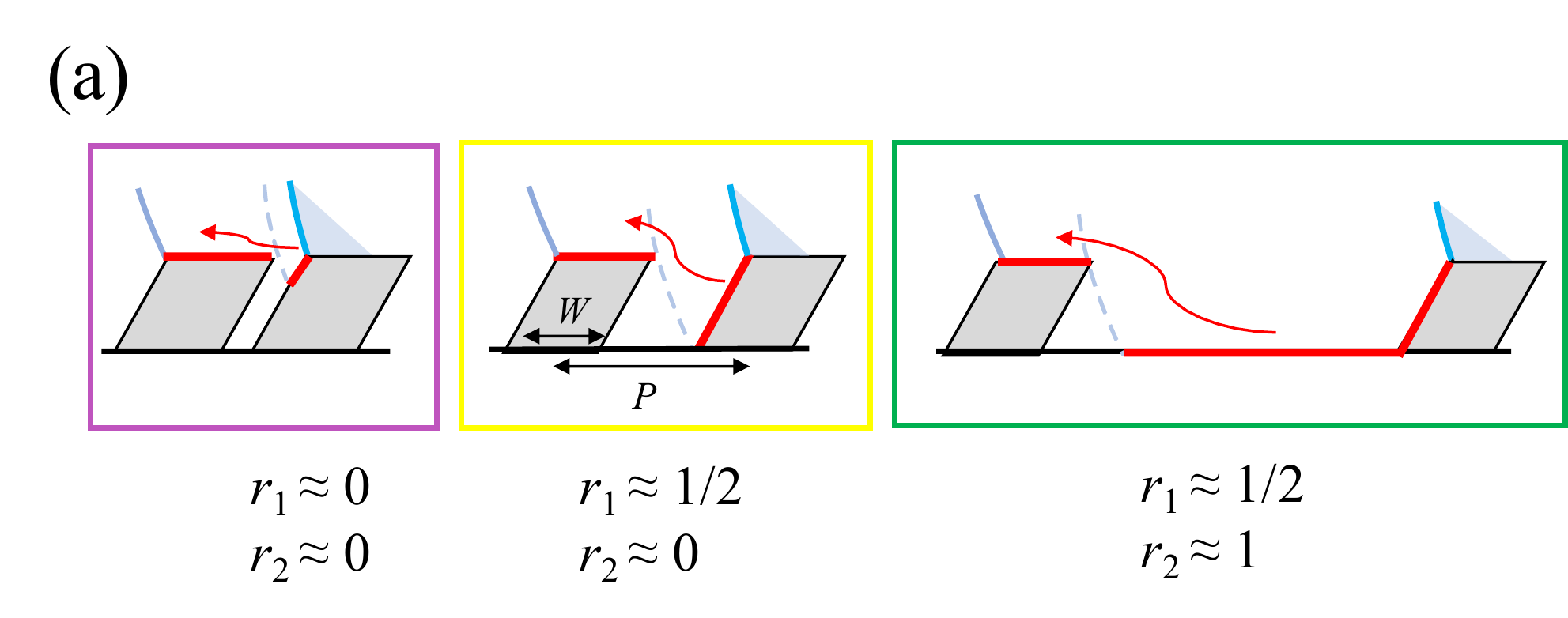}
  \includegraphics[width=0.35\textwidth]{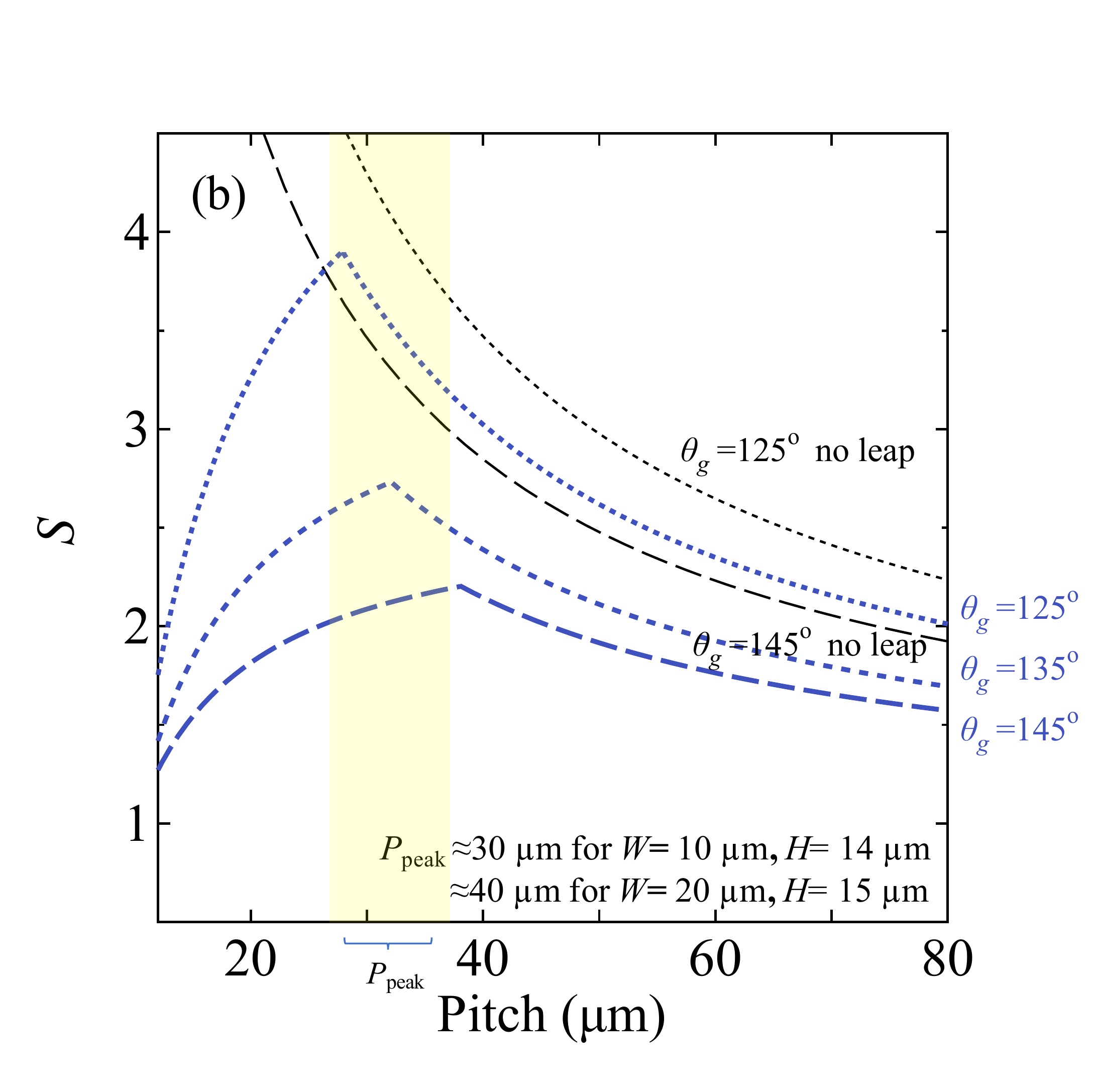}
  \includegraphics[width=0.33\textwidth]{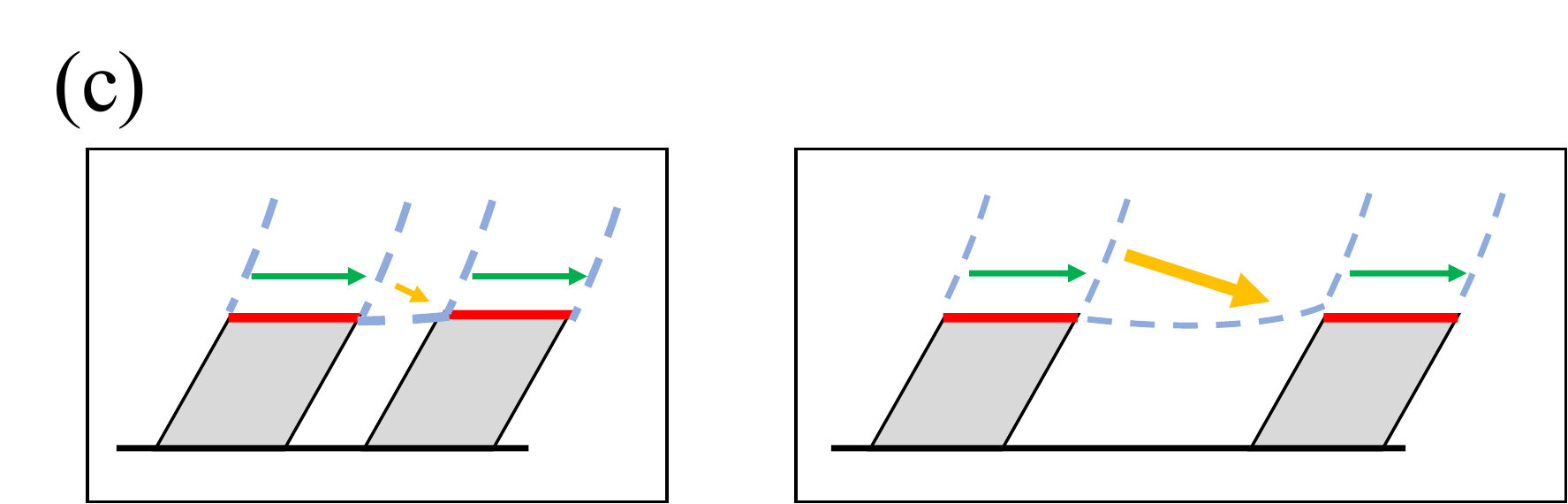}
  \includegraphics[width=0.35\textwidth]{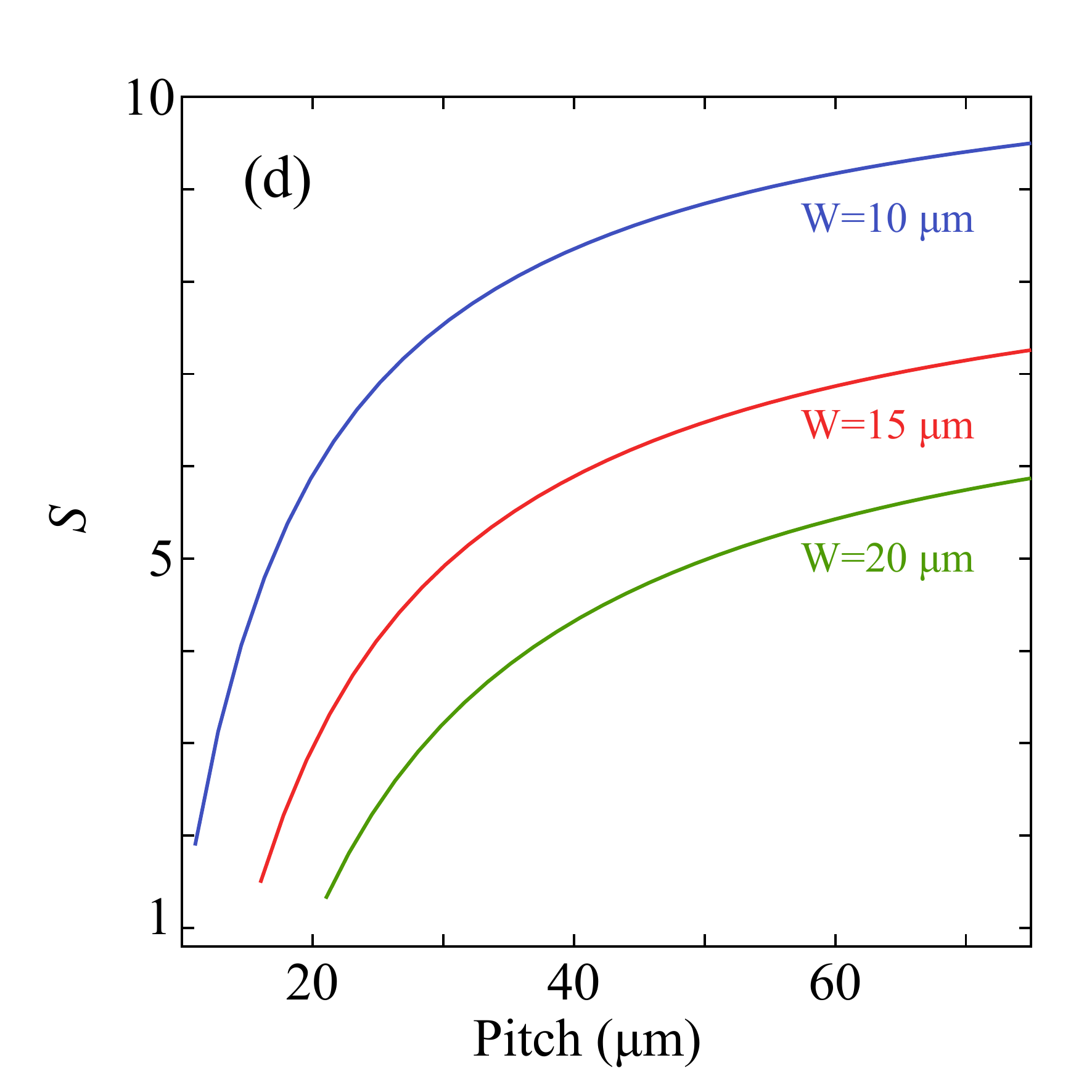}
  \end{center}
  \vspace{-0.2cm}
  \caption{
  (a) Sketches illustrating wetting situation in the direction against the inclination for small (left, $r_1\approx r_2\approx 0$), intermediate (center, $r_1\approx1/2, r_2\approx0$) and large (right, $r_1\approx 1/2,  r_2\approx 1$) pitch. Wetted sections are colored with red.
  (b) The passage time ratio $S$ in the direction against the inclination based on Eq.~(\ref{Eq:S_ag}) for $W=10$ $\mu$m as a function of the surface pitch. The blue lines correspond to $S$ obtain from the theoretical model for different apparent dynamic contact angles ($\theta_g=125^\circ, 135^\circ, 145^\circ$). The black line corresponds to passage times $S$ that would exists without leaping, i.e. assuming that the structured surface is wetted everywhere. 
  Without leaping, spreading becomes very slow for small pitch, which contradicts experimental observations where leaping will significantly increase the spreading speed.
  (c) Sketches illustrating wetting situation in the direction with the inclination for small(left) and large (right) pitch. 
  (d) The passage time ratio $S$ in the direction with the inclination based on Eq.~(\ref{eq:Sw}) for surface ridges of different widths as a function of the surface pitch. 
    }
\label{fig:model}
\end{figure}

\subsection{Model and scaling analysis} 
\subsubsection{A model for spread-and-leap regime}
In order to understand and to predict the observed wetting behaviour, we build simple models based on estimates of the relevant forces.

The non-monotonic behavior in Fig.~\ref{fig:pitch}(b) for W=20 $\mu$m can be explained by characterizing leaping over the microstructures in more detail. For a smooth flat surface, the passage time over a distance $P$ can be estimated as $\Tf \sim P/\Uf$, where $\Uf$ is the average contact-line velocity. For a surface with ridges, the wetted distance $W_s$ per period can be estimated as
$\Ws \approx 2r_1 H + r_2 P+ (1-r_2)W$.
Here, $r_1$ and $r_2$ represent the wetted portions of the vertical walls of the ridge and the valley between the ridges, respectively (see Fig.~\ref{fig:model}a). The time it takes for the contact line to move a distance $P$ is $\Ts \sim \Ws/\Us$, where $\Us$ is the pitch-averaged contact-line velocity. 
Inserting the scaling estimates for $\Ts$ and $\Tf$ in Eq.~(\ref{eq:Sa}), we obtain,
\begin{equation}
S \approx \frac{\Uf}{\Us}\left (2r_1\frac{H}{P} + r_2+ (1-r_2)\frac{W}{P}\right).
\label{eq:S2}
\end{equation}
For a very large pitch (rightmost frame in Fig.~\ref{fig:model}a), nearly all the structured surface is wetted ($r_1\approx1$ and $r_2\approx 1$); we have very small leaping and $S\sim 1+2H/P\approx 1$.  For a very small pitch (leftmost frame in Fig.~\ref{fig:model}a) the entire valleys between ridges remain dry ($r_1\approx 0$ and $r_2\approx 0$); we have maximum leaping and $S\sim W/P\approx 1$.  For intermediate values of the pitch $P$, where $r_1\approx 1/2$ and $r_2\approx 1/2$, a maximum value of the travel-time ratio $S$ exists.

To be more quantitative, we consider local contact line velocity based on phase field theory. As discussed in Section \ref{sec:regime}, the spreading in the direction against the inclination is in spread-and-leap regime. In this regime, we assume Young's force is driving the contact line and the line friction is the resistive force. 
Assuming that wetting resistance for spread-and-leap is dominated by line friction, we can develop a theoretical model of the contact-line velocity based on the Navier-Stokes-Cahn-Hilliard equations\cite{Yue2011,Lee2019},
\begin{equation}
U_{cl,i} = \frac{\sigma}{\mu_f}\frac{3}{2\sqrt{2}} \frac{\cos\theta_e-\cos\tl}{\sin\tl}.
\label{eq:cl}
\end{equation}
The wetted part of ridged surface is divided into $i=1,\dots, N$ smooth sections. Here, $\tl$ in Eq.~(\ref{eq:cl}) is the local dynamic contact angle formed between the liquid interface and the $i$th section of the structured surface.
We assume that Eq.~(\ref{eq:cl}) is a valid model of liquid front at any surface point \cite{Yue2011, Lee2019}. Considering the local dynamic contact angle and velocity, we sum up the time to pass sections $i=1,\dots,N$ 
\begin{equation}
	T_{s} = \sum_{i=1}^N{L_i/U_{cl,i}} \\
	=\sum_{i}{\frac{2\sqrt{2} \mu_{f}}{3 \sigma}\frac{L_{i}{\rm sin}\theta_{l,i}} {{\rm cos}\theta_{e}-{\rm cos}\theta_{l,i}}},
	\label{eq:Sl}
\end{equation}
and
\begin{equation}
\begin{aligned}
	S & = T_{s}/T_{flat} \\
	 & = \sum_{i}{(\frac{L_{i}{\rm sin}\theta_{l,i}} {{\rm cos}\theta_{eq}-{\rm cos}\theta_{l,i}})}\frac{{\rm cos}\theta_{eq}-{\rm cos}\theta_{g}}{L{\rm sin}\theta_{g}}, 	
	 \label{Eq:S_ag}
\end{aligned}
\end{equation}
where $\tl$, $U_{cl,i}$ and $L_i$ are the local dynamic contact angle, contact line velocity, and length of section $i$ respectively, and $\theta_g$ is the global apparent dynamic contact angle. The above effective spreading model is similar to that used in Lee et al.\cite{Lee2019}.  Compared to the sawtooth geometries analyzed by Lee et. al.\cite{Lee2019}., the geometry studied here is more complicated as it includes backward facing sections and sharper angles.  The present model accounts for the backward facing part of the structure. The droplet interface is assumed to be linear with a constant angle $\theta_g$, and the surface after the leap point is assumed to be not wetted. The wetted area is colored red in Fig.~\ref{fig:model}(a).

Figure \ref{fig:model}(b) shows the dependence of $S$ on the pitch obtained from the model for three different local dynamic contact angles. We observe that the model captures the trend of the resistance due to surface texture relatively well, confirming that line-friction is the dominating physical resistance for the spreading.
The spread-and-leap model explains the -- perhaps counter-intuitive -- fast spreading for small pitch. Without taking into account the leaping, a spreading model where all solid is wetted will overestimate $S$ significantly (Fig.~\ref{fig:model}b).
The model also predicts the pitch that gives the maximum $S$, $P_{peak}$. For W= 10 $\mu$m, $P_{peak}$ is 32 $\mu$m, 39 $\mu$m for W=15 $\mu$m, and 43 $\mu$m for W=20 $\mu$m (obtained assuming $\theta_g=135^\circ$). The estimate of $P_{peak}$ explains why the non-monotonic behaviour is observed only for W=20 $\mu$m. For W=10 $\mu$m and 15 $\mu$m, each $P$ is below $P_{peak}$. However for W=20 $\mu$m, $P_{peak}$ is close to the intermediate pitch 40 $\mu$m, therefore the non-monotonic behavior is observed.

\subsection{Scaling analysis for stick-and-leap regime}
The spreading in the direction with the inclination is in stick-and-leap regime. The contact line travels on the tip of the ridges and is pinned on the corner intermittently, and the interface bulges by the liquid inertia until it makes a new contact to the next structure. 
To understand the role of pinning, we perform a simple scaling analysis. In the direction with the inclination, Young's force is balanced both by inertia and line friction. Here, Young's force accelerates the spreading motion when the contact line is pulled over the structure (W2 $\rightarrow$ W3 in Fig.~\ref{fig:spread}f), and then adds to the droplet inertia. When the contact line reaches the acute corner at W3, it is momentarily pinned there.
The contact line remains pinned until the inertia in the droplet brings the local dynamic contact angle out of equilibrium such that the droplet interface can make contact with the solid in the forward direction. 

Assuming the surface energy is distributed to the kinetic energy over a whole droplet during the spreading on the top of the ridge, we have
\begin{equation}
	2 \pi R_0 \sigma W \sim \frac{4}{3} \pi \rho {R_{0}}^3 U^2,
\end{equation}	
and 
\begin{equation}
	U \sim \frac{\sqrt{3}}{R_{0}} \sqrt{\frac{\sigma W}{\rho}},
	\label{eq:Uinertial} 
\end{equation}	
 where $R_0$ is the initial radius of the droplet, $W$ is the width of the ridge, $L$ is the space between the ridges, and $U$ is a characteristic velocity for inertia. The prefactor $\sqrt{3}$ in Eq.~(\ref{eq:Uinertial})  can be neglected for simplicity.

Based on this physical insight as depicted in Fig.~\ref{fig:model}(c), we can model the passage time for stick-and-leap as 
\begin{equation}
    \Ts \sim \frac{\mu_f W}{\sigma}+\frac{(P-W) R_0}{\sqrt{\sigma W/\rho}}.
\label{eq:TSW}
\end{equation}
Here, the first term quantifies the balance between Young's force and line friction, i.e., the travel time for the interface to move over the top of the ridge of width $W$. 
The second term quantifies the pinning time, i.e., the time needed for inertia to push the interface to the next ridge, leaping the distance $P-W$.
Recalling that $\Tf \sim P/\Uf \approx P\mu_f/\sigma$, we can formulate the passage time $S$ as 
\begin{equation}
    S = \frac{W}{P}+\frac{1}{Oh_f}\sqrt{\frac{R_0}{W}}\frac{(P-W)}
  {P} .
  \label{eq:Sw}
\end{equation}
 We observe that the second term  dominates when inertia is much larger than the line friction ($Oh_f\ll 1$). 
For $P\gg W$, $S$ in Eq.~(\ref{eq:Sw}) saturates to the  constant value $1/Oh_f \sqrt{R_0/W}$. In our scaling analysis, we have assumed that this saturation value is reached when the liquid-gas interface remains pinned at the acute corner infinitely long. 

Figure \ref{fig:model}(d) presents $S$ obtained from Eq.~(\ref{eq:Sw}) for different ridge widths. We observe that the scaling estimates for stick-and-slip correctly capture the trend of increasing passage time with increasing pitch.  The scaling analysis shows that inertia plays a central role in the stick-and-leap regime, in contrast to spread-and-leap in the direction against inclination. Surface energy is converted to the kinetic energy of the droplet ($\sim$ inertia) while the contact line spreads on the top of the ridge, and inertia drives the liquid front to reach the next rise of the surface during the stick-and-leap at corners (W2 in Fig.~\ref{fig:spread}f).

We note that the advancement of the contact line is primarily determined by the line friction and the geometrical details of the structure when moving against the inclination, but that it is inertial when moving with the inclination. 

\vspace{0.4cm}
\section{Conclusions}
We have presented a comprehensive study of rapid wetting on complex asymmetric microstructures. 
 We identified three wetting regimes, denoted as spread, stick, and leap. 
By coordinated simulations and experiments we can follow the passage of the contact line over 
the microstructure in detail and formulate a model as well as scaling estimates that explain how the geometrical features determine the macroscopic wetting speed.
We showed that when wetting proceeds against the inclination of the ridges, a spread-and-leap behavior underpins the wetting and the driving Young's force is primarily balanced by contact line friction. In contrast, when the spreading direction is with the inclination, a stick-and-leap behavior is observed, and it is the liquid inertia that limits the wetting speed. Our experiments and theory show that leaping phenomenon plays a central role in increasing the spreading speed compared to a surface textured without leaping. We believe that this newly identified spreading mechanism forms the foundation to design surface structures for controlling wetting under realistic unsteady environments.

\section*{Conflicts of interest}
There are no conflicts to declare.

\section*{Acknowledgements}
This work was supported by the Swedish Research Council (VR 2015-04019). S.B acknowledges the support of the Swedish Foundation of Strategic research (SSF-FFL6).
We thank Calvin J. Brett at KTH for his kind help for SEM measurements.  



\balance


\bibliographystyle{rsc} 

\bibliography{rsc.bib} 

\end{document}